# Performance of a Relay-Assisted Hybrid FSO / RF Communication System

*Mohammad Ali Amirabadi*[1,✉]

[1] *School of Electrical Engineering, Iran University of Science and Technology, Tehran, Iran*
✉ E-mail: m_amirabadi@elec.iust.ac.ir

**Abstract:** This paper investigates performance of a relay assisted hybrid Free Space Optical / Radio Frequency (FSO/RF) communication system. The proposed structure is particularly recommended anywhere that direct RF communication between mobile users and base station is not possible due to atmospheric conditions. In this system, a multiuser RF link connects mobile users to relay and a FSO link connects relay to the base station. It is the first time that effect of number of users within the cell, on the performance of such structure, is investigated. Also it is the first time that performance of a dual hop hybrid FSO / RF system is investigated in Negative Exponential atmospheric turbulence. Considering wide range of atmospheric turbulence regimes, from moderate to saturated, for the first time, closed-form expressions are derived for Bit Error Rate (BER) and outage probability ($P_{out}$) of the proposed structure. MATLAB simulations verified accuracy of the derived expressions. Considering fixed and adaptive gain amplify and forward protocols at relay, it is shown that adaptive gain relay is less sensitive to the number of users within the cell. It is also shown that fixed gain relay, despite its low complexity, has better performance; because its gain is adjusted such that the performance be favorable even at the worst case scenario.

## 1 Introduction

Recently FSO communication system is taken into consideration in research studies. This is because of its high data rate, bandwidth and security, in an unlicensed spectrum as well as easy and low cost installation [1][2]. In contrast with these advantages, high sensitivity of FSO system to weather conditions and atmospheric turbulences limits its practical applications.

Scintillation is one of the effects of atmospheric turbulences on the performance of FSO system. It causes intensity fluctuation of the received signal. To reduce this effect, aperture averaging at the receiver can be used [3]. In this technique, receiver aperture diameter changes adaptively based on Channel State Information (CSI). Another factor that degrades performance of FSO system is pointing error. It is due to the non-linearity of transmitter and receiver, caused by building sway in the wind, small earthquakes, and building vibrations. One way to deal with this effect is to increase the beam diameter [4]. Various statistical distributions have been used in papers to model this effects. Mostly, Lognormal distribution is used to model weak atmospheric turbulence, Gamma – Gamma distribution is used to model moderate to strong atmospheric turbulence and Negative Exponential distribution is used to model saturate atmospheric turbulence [5].

Impact of atmospheric conditions on FSO and RF links is not the same; e. g. in FSO link the main performance degradation factors are fog and atmospheric turbulences, and the rain does not have much impact on it; while, RF link is sensitive to heavy rain but does not care about atmospheric turbulence and fog [6].

One way to improve performance of both FSO and RF links is to combine them together. In fact, FSO and RF have complementary aspects [7]. Hybrid FSO / RF systems are available in parallel and series structures. In parallel structure, millimeter wave RF link acts as a backup of FSO link. Series structures are relay assisted [4], and in mobile infrastructure, RF and FSO links respectively connect users to relay and relay to base station. Therefore, series structures are suitable for areas where direct RF connection between users and base station is impossible.

Relay assisted hybrid FSO/RF system, has advantages of FSO, RF and relay assisted systems, such as high bandwidth, more reliability and better performance, all at the same time. Different protocols are used at relay, one of the mostly used is amplify and forward [4]. In this protocol relay amplifies the received signal with fixed or adaptive gain based on channel conditions and forwards it.

FSO systems use various modulation schemes, one of the widely used modulations is On – Off keying (OOK). When this modulation is used, detection must be done according to a threshold based on CSI, so channel estimation is required. Another modulation is pulse position modulation (PPM) that does not require adaptive threshold detection, but has less spectral efficiency than OOK. Subcarrier intensity modulation (SIM), due to higher spectral efficiency over PPM and OOK, is an appropriate alternative for PPM and OOK, but requires carrier phase and frequency synchronization [8].

Several papers have studied relay assisted single hop [9] and dual hop [10] [11] structures. To the best of authors' knowledge, this is the first time that in a dual hop hybrid FSO / RF system, selection of the user with maximum signal to noise ratio (SNR) is considered in theoretical analysis. Also it is the first time that the performance of such structure is investigated over a wide range of atmospheric turbulence regimes from moderate to saturate. This paper for the first time investigates the effect of number of users within the cell on the performance of a dual hop hybrid FSO / RF system. FSO link, in moderate to strong atmospheric turbulence regimes, has Gamma - Gamma distribution with the effect of pointing errors, in saturated regime has Negative Exponential distribution and RF link has Rayleigh distribution. For the first time closed form expressions are derived for BER and $P_{out}$ of the proposed system. At the FSO receiver, Direct Detection / Intensity Modulation (IM /DD) is used.

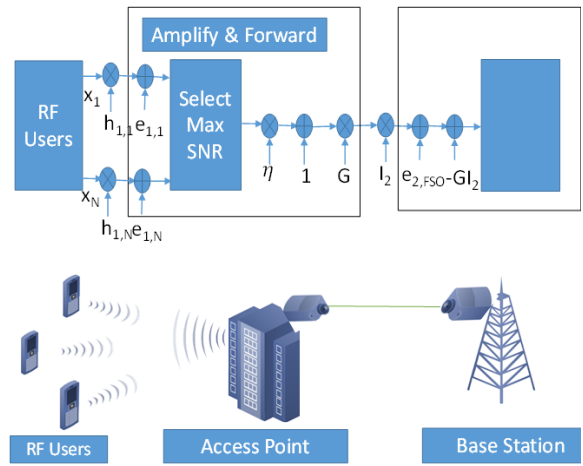

**Fig. 1:** Proposed relay assisted Hybrid FSO / RF system

This work was published for case of adaptive gain relaying in Gamma-Gamma atmospheric turbulence by the authors; this work is now extended to both cases of adaptive and fixed gain relaying at moderate to saturated atmospheric turbulence [12]. Rest of the paper is organized as follows: In Section II, the proposed relay assisted hybrid FSO / RF system model is presented. Section III studies performance of fixed gain scheme and section IV studies performance of adaptive gain scheme. Section V dedicates to comparison of analytic and simulation results, and section VI is conclusion of this paper.

## 2 System Model

In Fig. 1, the proposed dual hop relay assisted hybrid FSO/RF system is considered in which a relay is connected to mobile users via RF link and to base station via FSO link. Relay converts received RF signal to FSO signal with conversion efficiency of $\eta$, and a DC signal with unit amplitude is added to the FSO signal in order to be positive. Considering $x_i$; i = 1, 2, ..., N as the transmitted signal from i-th mobile user in RF link, the received signal at the relay is as follows:

$$y_{RF,i} = h_i x_i + e_{RF,i} \qquad (1)$$

where $h_i$, fading coefficient of the path between i-th user and relay, is complex Gaussian random variable [11]. $e_{RF,i}$, relay input noise, is Additive White Gaussian Noise (AWGN) with zero mean and $\sigma_{RF}^2$ variance. Relay connects with user that has highest SNR at relay input, i.e.:

$$\gamma_{RF} = max(\gamma_{RF,1}, \gamma_{RF,2}, ..., \gamma_{RF,N}) \qquad (2)$$

Relay converts received RF signal to FSO signal, then amplifies and forwards it to the base station as follows:

$$x_R = G(1 + \eta y_{RF}) \qquad (3)$$

where $G$ is the amplification gain and $y_{RF}$ is selected RF signal. After DC removal, received FSO signal becomes as follows:

$$y_{FSO} = I_2 x_R + e_{FSO} - I_2 G = I_2 G \eta (h x_i + e_{RF}) + e_{FSO} \qquad (4)$$

where $I_2$ is atmospheric turbulence intensity and $e_{FSO}$ is base station input noise, AWGN with $\sigma_{FSO}^2$ variance and zero mean. Instantaneous SNR at base station becomes as [13]:

$$\gamma_{FSO/RF} = \frac{I_2^2 G^2 \eta^2 h^2}{I_2^2 G^2 \eta^2 \sigma_{RF}^2 + \sigma_{FSO}^2} = \frac{\frac{I_2^2 \eta^2}{\sigma_{FSO}^2} \frac{h^2}{\sigma_{RF}^2}}{\frac{I_2^2 \eta^2}{\sigma_{FSO}^2} + \frac{1}{G^2 \sigma_{RF}^2}} \qquad (5)$$

where the signal power is assumed to be unit ($E[x_i^2] = 1$). In the case of CSI non-existence, received signal is amplified with fixed gain $G^2 = 1/(C\sigma_{RF}^2)$ [13]; where $C$ is a desired constant parameter. By substitution $G$ and $\gamma_{RF} = h^2/\sigma_{RF}^2$ and $\gamma_{FSO} = \eta^2 I_2^2/\sigma_{FSO}^2$ in (5), instantaneous SNR at base station input for fixed gain scheme becomes as:

$$\gamma_{FSO/RF} = \frac{\gamma_{FSO}\gamma_{RF}}{\gamma_{FSO}+C} \qquad (6)$$

As can be seen, if $\gamma_{RF}$ and $\gamma_{FSO}$ tend to infinity, then $\gamma_{FSO/RF} \cong \gamma_{RF}$. In the sense that in fixed gain scheme, at high SNRs, system performance is independent of FSO link atmospheric turbulence. When CSI exists at the relay, received signal would be amplified with adaptive gain as $G^2 = 1/(h^2 + \sigma_{RF}^2)$ [13]. As can be seen, adaptive gain scheme needs channel estimation but fixed gain scheme does not need it. Substituting $G$ and $\gamma_{RF} = h^2/\sigma_{RF}^2$ and $\gamma_{FSO} = \eta^2 I_2^2/\sigma_{FSO}^2$ in (5), instantaneous SNR at base station input for adaptive gain scheme becomes:

$$\gamma_{FSO/RF} = \frac{\gamma_{FSO}\gamma_{RF}}{\gamma_{FSO}+\gamma_{RF}+1} \qquad (7)$$

If $\gamma_{RF}$, $\gamma_{FSO}$ tend to infinity, then $\gamma_{FSO/RF} \cong min(\gamma_{FSO}, \gamma_{RF})$; in the sense that at high SNR, link with lower SNR level, has the most impact on performance of the system.

Probability density function (pdf) of Gamma-Gamma distribution with the effect of pointing errors is as [14]:

$$f_{\gamma_{FSO}}(\gamma) = \frac{\xi^2}{2\Gamma(\alpha)\Gamma(\beta)\gamma} G_{1,3}^{3,0}\left(\alpha\beta\kappa\sqrt{\frac{\gamma}{\bar{\gamma}_{FSO}}}\Big|_{\xi^2,\alpha,\beta}^{\xi^2+1}\right) \qquad (8)$$

where $G_{p,q}^{m,n}\left(z\Big|_{b_1,b_2,...,b_q}^{a_1,a_2,...,a_p}\right)$ is Meijer-G function, $\alpha, \beta$ are parameters related to Gamma-Gamma atmospheric turbulence intensity, $\xi^2$ is related to the effect of pointing errors and $\Gamma(.)$ is Gamma function [15]. $\xi^2 = \omega_{Z_{eq}}/(2\sigma_s)$ is ratio between equivalent beam diameter and pointing error displacement standard deviation (jitter) at the receiver. Where $\sigma_s^2$ is the jitter variance and $\omega_{Z_{eq}}$ is equivalent beam radius at the receiver. $\alpha, \beta$ are defined as $\alpha = \left[\exp\left(0.49\sigma_R^2/(1 + 1.11\sigma_R^{\frac{12}{5}})^{\frac{7}{6}}\right) - 1\right]^{-1}$ and $\beta = \left[\exp\left(0.51\sigma_R^2/(1 + 0.69\sigma_R^{\frac{12}{5}})^{\frac{5}{6}}\right) - 1\right]^{-1}$, where $\sigma_R^2$ is the Rytov variance [14]. Average SNR at the FSO receiver input is $\bar{\gamma}_{FSO} = 1/\sigma_{FSO}^2$.

Cumulative Distribution Function (CDF) of Gamma-Gamma distribution with the effect of pointing error is as follows [14]:

$$F_{\gamma_{FSO}}(\gamma) = \frac{\xi^2}{\Gamma(\alpha)\Gamma(\beta)} G_{2,4}^{3,1}\left(\alpha\beta\kappa\sqrt{\frac{\gamma}{\bar{\gamma}_{FSO}}}\Big|_{\xi^2,\alpha,\beta,0}^{1,\xi^2+1}\right) \qquad (9)$$

The pdf and CDF of Negative Exponential distribution with $1/\lambda^2$ variance and $1/\lambda$ mean are respectively as follows:

$$f_{\gamma_{FSO}}(\gamma) = \frac{\lambda}{2\sqrt{\gamma\bar{\gamma}_{FSO}}} e^{-\lambda\sqrt{\frac{\gamma}{\bar{\gamma}_{FSO}}}} \qquad (10)$$

$$F_{\gamma_{FSO}}(\gamma) = 1 - e^{-\lambda\sqrt{\frac{\gamma}{\bar{\gamma}_{FSO}}}} \qquad (11)$$

Fading distribution of i-th RF path is Rayleigh and its variance is assumed to be the same in all paths ($\bar{\gamma}_{RF,i} = \bar{\gamma}_{RF}$). The pdf of $\gamma_{RF,i}$ random variable is as follows [13]:

$$f_{\gamma_{RF,i}}(\gamma) = \frac{1}{\bar{\gamma}_{RF}} e^{-\frac{\gamma}{\bar{\gamma}_{RF}}} \quad (12)$$

Integrating the above equation, CDF of $\gamma_{RF,i}$ random variable becomes:

$$F_{\gamma_{RF,i}}(\gamma) = 1 - e^{-\frac{\gamma}{\bar{\gamma}_{RF}}} \quad (13)$$

Using (2), CDF of the entire RF link is as follows:

$$F_{\gamma_{RF}}(\gamma) = Pr(max(\gamma_{RF,1}, \gamma_{RF,2}, \ldots, \gamma_{RF,N}) \leq \gamma) = Pr(\gamma_{RF,1} \leq \gamma, \gamma_{RF,2} \leq \gamma, \ldots, \gamma_{RF,N} \leq \gamma) \quad (14)$$

Assuming RF paths independence, CDF of entire RF link becomes:

$$F_{\gamma_{RF}}(\gamma) = \prod_{i=1}^{N} Pr(\gamma_{RF,i} \leq \gamma) = \prod_{i=1}^{N} F_{\gamma_{RF,i}}(\gamma) \quad (15)$$

Assuming identity RF fades, CDF of entire RF link becomes:

$$F_{\gamma_{RF}}(\gamma) = \left(F_{\gamma_{RF,i}}(\gamma)\right)^N = \left(1 - e^{-\frac{\gamma}{\bar{\gamma}_{RF}}}\right)^N \quad (16)$$

By derivation of (16), pdf of entire RF link becomes:

$$f_{\gamma_{RF}}(\gamma) = N\left(F_{\gamma_{RF,i}}(\gamma)\right)^{N-1} f_{\gamma_{RF,i}}(\gamma) = \frac{N}{\bar{\gamma}_{RF}} e^{-\frac{\gamma}{\bar{\gamma}_{RF}}} \left(1 - e^{-\frac{\gamma}{\bar{\gamma}_{RF}}}\right)^{N-1} \quad (17)$$

Substituting binomial expansion of $\left(1 - e^{-\gamma/\bar{\gamma}_{RF}}\right)^{N-1}$ as $\sum_{k=0}^{N-1} \binom{N-1}{k}(-1)^k e^{-k\gamma/\bar{\gamma}_{RF}}$ into (17), pdf of entire RF link becomes:

$$f_{\gamma_{RF}}(\gamma) = \frac{N}{\bar{\gamma}_{RF}} \sum_{k=0}^{N-1} \binom{N-1}{k}(-1)^k e^{-\frac{(k+1)\gamma}{\bar{\gamma}_{RF}}} \quad (18)$$

## 3 Performance of fixed gain structure

### 3.1. Outage Probability

When instantaneous SNR reduces below a threshold level ($\gamma_{th}$), outage occurs. According to this definition, $P_{out}$ of the proposed system is as follows:

$$P_{out,\gamma_{FSO/RF}}(\gamma_{th}) = Pr(\gamma_{FSO/RF} \leq \gamma_{th}) = 1 - Pr\left(\gamma_{\frac{FSO}{RF}} \geq \gamma_{th}\right) = 1 - Pr\left(\frac{\gamma_{FSO}\gamma_{RF}}{\gamma_{FSO}+C} \geq \gamma_{th}\right) \quad (19)$$

After mathematical simplification, the above equation becomes as follows [17]:

$$= 1 - \int_0^\infty Pr\left(\gamma_{FSO} \geq \frac{\gamma_{th}C}{x}|\gamma_{RF}\right) f_{\gamma_{RF}}(x + \gamma_{th}) dx \quad (20)$$

Substituting (9) and (18) into (20) and after mathematical simplification, $P_{out}$ of the proposed system in Gamma-Gamma atmospheric turbulence considering pointing error becomes as follows:

$$P_{out,\gamma_{FSO/RF}}(\gamma_{th}) = 1 - \frac{N}{\bar{\gamma}_{RF}} \sum_{k=0}^{N-1} \binom{N-1}{k}(-1)^k e^{-\frac{(k+1)\gamma_{th}}{\bar{\gamma}_{RF}}} \left[\int_0^\infty e^{-\frac{(k+1)x}{\bar{\gamma}_{RF}}} dx - \frac{\xi^2}{\Gamma(\alpha)\Gamma(\beta)} \int_0^\infty e^{-\frac{(k+1)x}{\bar{\gamma}_{RF}}} G_{2,4}^{3,1}\left(\alpha\beta\kappa\sqrt{\frac{\gamma_{th}C}{x\bar{\gamma}_{FSO}}}\Big|_{\xi^2,\alpha,\beta,0}^{1,\xi^2+1}\right) dx\right] \quad (21)$$

Substituting equivalent of $G_{2,4}^{3,1}\left(\alpha\beta\kappa\sqrt{\frac{\gamma_{th}C}{x\bar{\gamma}_{FSO}}}\Big|_{\xi^2,\alpha,\beta,0}^{1,\xi^2+1}\right)$ as $G_{4,2}^{1,3}\left(\frac{1}{\alpha\beta\kappa}\sqrt{\frac{x\bar{\gamma}_{FSO}}{\gamma_{th}C}}\Big|_{0,-\xi^2}^{1-\xi^2,1-\alpha,1-\beta,1}\right)$ [16,Eq. 07.34.17.0012.01] and by using [16,Eq. 07.34.21.0088.01], $P_{out}$ of the proposed system in Gamma-Gamma atmospheric turbulence with the effect of pointing errors becomes:

$$P_{out,\gamma_{FSO/RF}}(\gamma_{th}) = 1 - \sum_{k=0}^{N-1}\binom{N-1}{k}(-1)^k \frac{N}{k+1} e^{-\frac{(k+1)\gamma_{th}}{\bar{\gamma}_{RF}}} \left[1 - \frac{\xi^2 2^{\alpha+\beta-3}}{\pi\Gamma(\alpha)\Gamma(\beta)} G_{9,4}^{2,7}\left(\frac{16\bar{\gamma}_{FSO}\bar{\gamma}_{RF}}{(\alpha\beta\kappa)^2\gamma_{th}C(k+1)}\Big|_{0,\frac{1}{2},\frac{\xi^2}{2},\frac{1-\xi^2}{2}}^{0,\frac{1-\xi^2}{2},\frac{2-\xi^2}{2},\frac{1-\alpha}{2},\frac{2-\alpha}{2},\frac{1-\beta}{2},\frac{2-\beta}{2},\frac{1}{2},1}\right)\right] \quad (22)$$

Substituting (11) and (18) into (20) and after mathematical simplification, $P_{out}$ of proposed system in Negative Exponential atmospheric turbulence is equal to:

$$P_{out,\gamma_{FSO/RF}}(\gamma_{th}) = 1 - \frac{N}{\bar{\gamma}_{RF}} \sum_{k=0}^{N-1} \binom{N-1}{k}(-1)^k e^{-\frac{(k+1)\gamma_{th}}{\bar{\gamma}_{RF}}} \int_0^\infty e^{-\frac{(k+1)x}{\bar{\gamma}_{RF}}} e^{-\lambda\sqrt{\frac{\gamma_{th}C}{x\bar{\gamma}_{FSO}}}} dx \quad (23)$$

From [16, Eq.07.34.17.0012.01] and [16, Eq. 07.34.03.1081.01] Meijer-G equivalent of $e^{-\lambda\sqrt{C\gamma_{th}/(x\bar{\gamma}_{FSO})}}$ is equal to $\frac{1}{\sqrt{\pi}} G_{2,0}^{0,2}\left(4x\bar{\gamma}_{FSO}/(\lambda^2\gamma_{th}C)\Big|^{1,1/2}\right)$. Substituting it and using [16, Eq. 07.34.21.0088.01] $P_{out}$ of the proposed system in Negative Exponential atmospheric turbulence is equal to:

$$P_{out,\gamma_{FSO/RF}}(\gamma_{th}) = 1 - \sum_{k=0}^{N-1}\binom{N-1}{k}(-1)^k \frac{N}{\sqrt{\pi}(k+1)} e^{-\frac{(k+1)\gamma_{th}}{\bar{\gamma}_{RF}}} \times G_{3,0}^{0,3}\left(\frac{4\bar{\gamma}_{FSO}\bar{\gamma}_{RF}}{\lambda^2\gamma_{th}C(k+1)}\Big|_{\frac{1}{2}}^{0,1,1}\right) \quad (24)$$

### 3.2. Bit Error Rate

Although MPSK modulations have better BER performance than DPSK, but DPSK does not require carrier phase estimation circuit and has low complexity receiver. Given that $F_\gamma(\gamma) = P_{out}(\gamma)$, BER of DPSK modulation can be obtained from the following equation [13]:

$$P_e = \frac{1}{2}\int_0^\infty e^{-\gamma} F_\gamma(\gamma) d\gamma = \frac{1}{2}\int_0^\infty e^{-\gamma} P_{out}(\gamma) d\gamma \quad (25)$$

Substituting (22) into (25), BER of DPSK modulation over Gamma-Gamma atmospheric turbulence considering pointing error is equal to:

$$= \frac{1}{2}\int_0^\infty e^{-\gamma}\left\{1 - \sum_{k=0}^{N-1}\binom{N-1}{k}(-1)^k \frac{N}{k+1} e^{\frac{(k+1)\gamma}{\bar{\gamma}_{RF}}}\left[1 - \frac{\xi^2 2^{\alpha+\beta-3}}{\pi\Gamma(\alpha)\Gamma(\beta)} \times G_{9,4}^{2,7}\left(\frac{16\bar{\gamma}_{FSO}\bar{\gamma}_{RF}}{(\alpha\beta\kappa)^2\gamma C(k+1)}\middle| \begin{matrix}0,\frac{1-\xi^2}{2},\frac{2-\xi^2}{2},\frac{1-\alpha}{2},\frac{2-\alpha}{2},\frac{1-\beta}{2},\frac{2-\beta}{2},\frac{1}{2},1\\ 0,\frac{1}{2},\frac{-\xi^2}{2},\frac{1-\xi^2}{2}\end{matrix}\right)\right]\right\}d\gamma$$

(26)

Substituting equivalent of

$$G_{9,4}^{2,7}\left(\frac{16\bar{\gamma}_{FSO}\bar{\gamma}_{RF}}{(\alpha\beta\kappa)^2\gamma C(k+1)}\middle| \begin{matrix}0,\frac{1-\xi^2}{2},\frac{2-\xi^2}{2},\frac{1-\alpha}{2},\frac{2-\alpha}{2},\frac{1-\beta}{2},\frac{2-\beta}{2},\frac{1}{2},1\\ 0,\frac{1}{2},\frac{-\xi^2}{2},\frac{1-\xi^2}{2}\end{matrix}\right)$$ as

$$G_{4,9}^{7,2}\left(\frac{(\alpha\beta\kappa)^2\gamma C(k+1)}{16\bar{\gamma}_{FSO}\bar{\gamma}_{RF}}\middle| \begin{matrix}1,\frac{1}{2},\frac{2+\xi^2}{2},\frac{1+\xi^2}{2}\\ 1,\frac{1+\xi^2}{2},\frac{\xi^2}{2},\frac{1+\alpha}{2},\frac{\alpha}{2},\frac{1+\beta}{2},\frac{\beta}{2},\frac{1}{2},0\end{matrix}\right)^0$$ [16,

Eq.07.34.17.0012.01] and using [16, Eq. 07.34.21.0088.01] BER of DPSK modulation over Gamma-Gamma atmospheric turbulence with the effect of pointing error becomes equal to:

$$P_e = \frac{1}{2}\left\{1 - \sum_{k=0}^{N-1}\binom{N-1}{k}(-1)^k \frac{N}{k+1}\frac{1}{1+\frac{(k+1)}{\bar{\gamma}_{RF}}}\left[1 - \frac{\xi^2 2^{\alpha+\beta-3}}{\pi\Gamma(\alpha)\Gamma(\beta)} \times G_{5,9}^{7,3}\left(\frac{(\alpha\beta\kappa)^2 C(k+1)}{16\bar{\gamma}_{FSO}(\bar{\gamma}_{RF}+k+1)}\middle| \begin{matrix}0,1,\frac{1}{2},\frac{2+\xi^2}{2},\frac{1+\xi^2}{2}\\ 1,\frac{1+\xi^2}{2},\frac{\xi^2}{2},\frac{1+\alpha}{2},\frac{\alpha}{2},\frac{1+\beta}{2},\frac{\beta}{2},\frac{1}{2},0\end{matrix}\right)\right]\right\}$$

(27)

Substituting (24) into (25), BER of negative exponential atmospheric turbulence is equal to:

$$P_e = \frac{1}{2}\int_0^\infty e^{-\gamma}\left\{1 - \sum_{k=0}^{N-1}\binom{N-1}{k}(-1)^k \times \frac{N}{\sqrt{\pi}(k+1)} e^{\frac{(k+1)\gamma}{\bar{\gamma}_{RF}}} G_{3,0}^{0,3}\left(\frac{4\bar{\gamma}_{FSO}\bar{\gamma}_{RF}}{\lambda^2\gamma C(k+1)}\middle|\begin{matrix}0,1,\frac{1}{2}\\ -\end{matrix}\right)\right\}d\gamma \quad (28)$$

Substituting equivalent of $G_{3,0}^{0,3}\left(\frac{4\bar{\gamma}_{FSO}\bar{\gamma}_{RF}}{\lambda^2\gamma C(k+1)}\middle|\begin{matrix}0,1,0.5\\ -\end{matrix}\right)$ as $G_{0,3}^{3,0}\left(\frac{\lambda^2\gamma C(k+1)}{4\bar{\gamma}_{FSO}\bar{\gamma}_{RF}}\middle|\begin{matrix}-\\ 1,0,0.5\end{matrix}\right)$ [16, Eq.07.34.17.0012.01] and using [16, Eq.07.34.21.0088.01] BER of Negative Exponential atmospheric turbulence is equal to:

$$P_e = \frac{1}{2}\left\{1 - \sum_{k=0}^{N-1}\binom{N-1}{k}(-1)^k \frac{N}{\sqrt{\pi}(k+1)}\frac{1}{1+\frac{(k+1)}{\bar{\gamma}_{RF}}} \times G_{1,3}^{3,1}\left(\frac{4\bar{\gamma}_{FSO}\bar{\gamma}_{RF}}{\lambda^2\gamma C(k+1)}\middle|\begin{matrix}0\\ 1,0,\frac{1}{2}\end{matrix}\right)\right\} \quad (29)$$

## 4 Performance of Adaptive gain structure

### 4.1. Outage Probability

(7) can be approximated as [13]:

$$\gamma_{FSO/RF} = \frac{\gamma_{FSO}\gamma_{RF}}{\gamma_{FSO}+\gamma_{RF}+1} \cong \min(\gamma_{FSO},\gamma_{RF}) \quad (30)$$

CDF of $\gamma_{FSO/RF}$ random variable equals with [18]:

$$F_{\gamma_{FSO/RF}}(\gamma) = F_{\gamma_{RF}}(\gamma) + F_{\gamma_{FSO}}(\gamma) - F_{\gamma_{RF}}(\gamma)F_{\gamma_{FSO}}(\gamma) \quad (31)$$

Given that $P_{out}(\gamma_{th}) = F_\gamma(\gamma_{th})$, Substituting (9) and (16) into (31), $P_{out}$ of Gamma-Gamma atmospheric turbulence with the effect of pointing error is equal to:

$$P_{out,\gamma_{FSO/RF}}(\gamma_{th}) = \left(1 - e^{-\frac{\gamma_{th}}{\bar{\gamma}_{RF}}}\right)^N + \frac{\xi^2}{\Gamma(\alpha)\Gamma(\beta)}G_{2,4}^{3,1}\left(\alpha\beta\kappa\sqrt{\frac{\gamma_{th}}{\bar{\gamma}_{FSO}}}\middle|\begin{matrix}1,\xi^2+1\\ \xi^2,\alpha,\beta,0\end{matrix}\right) - \frac{\xi^2}{\Gamma(\alpha)\Gamma(\beta)}\left(1 - e^{-\frac{\gamma_{th}}{\bar{\gamma}_{RF}}}\right)^N G_{2,4}^{3,1}\left(\alpha\beta\kappa\sqrt{\frac{\gamma_{th}}{\bar{\gamma}_{FSO}}}\middle|\begin{matrix}1,\xi^2+1\\ \xi^2,\alpha,\beta,0\end{matrix}\right) \quad (32)$$

Given that $P_{out}(\gamma_{th}) = F_\gamma(\gamma_{th})$, by substituting (11) and (16) into (31), $P_{out}$ of Negative Exponential atmospheric turbulence is equal to:

$$P_{out,\gamma_{FSO/RF}}(\gamma_{th}) = \left(1 - e^{-\frac{\gamma_{th}}{\bar{\gamma}_{RF}}}\right)^N + \left(1 - e^{-\lambda\sqrt{\frac{\gamma_{th}}{\bar{\gamma}_{FSO}}}}\right) - \left(1 - e^{-\frac{\gamma_{th}}{\bar{\gamma}_{RF}}}\right)^N\left(1 - e^{-\lambda\sqrt{\frac{\gamma_{th}}{\bar{\gamma}_{FSO}}}}\right) \quad (33)$$

### 4.2. Bit Error Rate

By substituting (32) into (25), BER of DPSK modulation over gamma-gamma atmospheric turbulence considering pointing errors equals with:

$$P_e = \frac{1}{2}\int_0^\infty e^{-\gamma}\left\{\left(1 - e^{-\frac{\gamma}{\bar{\gamma}_{RF}}}\right)^N + \frac{\xi^2}{\Gamma(\alpha)\Gamma(\beta)}G_{2,4}^{3,1}\left(\alpha\beta\kappa\sqrt{\frac{\gamma}{\bar{\gamma}_{FSO}}}\middle|\begin{matrix}1,\xi^2+1\\ \xi^2,\alpha,\beta,0\end{matrix}\right) - \frac{\xi^2}{\Gamma(\alpha)\Gamma(\beta)}\left(1 - e^{-\frac{\gamma}{\bar{\gamma}_{RF}}}\right)^N G_{2,4}^{3,1}\left(\alpha\beta\kappa\sqrt{\frac{\gamma}{\bar{\gamma}_{FSO}}}\middle|\begin{matrix}1,\xi^2+1\\ \xi^2,\alpha,\beta,0\end{matrix}\right)\right\}d\gamma \quad (34)$$

Substituting the binomial expansion of $\left(1 - e^{-\gamma/\bar{\gamma}_{RF}}\right)^N$ as $\sum_{k=0}^N\binom{N}{k}(-1)^k e^{k\gamma/\bar{\gamma}_{RF}}$, and using [16, Eq. 07.34.21.0088.01] BER of DPSK modulation over Gamma-Gamma atmospheric turbulence with the effect of pointing errors equals with:

$$P_e = \frac{1}{2}\left\{\frac{\xi^2 2^{\alpha+\beta-3}}{\pi\Gamma(\alpha)\Gamma(\beta)}G_{5,8}^{6,3}\left(\frac{(\alpha\beta\kappa)^2}{16\bar{\gamma}_{FSO}}\middle|\begin{matrix}0,1,\frac{1+\xi^2}{2},\frac{2+\xi^2}{2}\\ \frac{\xi^2}{2},\frac{1+\xi^2}{2},\frac{1+\alpha}{2},\frac{\alpha}{2},\frac{1+\beta}{2},\frac{\beta}{2},0,\frac{1}{2}\end{matrix}\right)\right. + \quad (35)$$

$$\sum_{k=0}^{N}\binom{N}{k}(-1)^k \frac{1}{1+\frac{k}{\bar{\gamma}_{RF}}}\left[1-\right.$$
$$\left.\frac{\xi^2 2^{\alpha+\beta-3}}{\pi\Gamma(\alpha)\Gamma(\beta)}G_{5,8}^{6,3}\left(\frac{(\alpha\beta\kappa)^2}{16\bar{\gamma}_{FSO}\left(1+\frac{k}{\bar{\gamma}_{RF}}\right)}\Big|_{\frac{\xi^2}{2},\frac{1+\xi^2}{2},\frac{1+\alpha}{2},\frac{\alpha}{2},\frac{1+\beta}{2},\frac{\beta}{2},0,\frac{1}{2}}^{0,1,\frac{1+\xi^2}{2},\frac{2+\xi^2}{2}}\right)\right]\Big\}$$

Substituting (33) into (25), BER of DPSK modulation over Negative Exponential atmospheric turbulence equals with:

$$P_e = \frac{1}{2}\int_0^\infty e^{-\gamma}\left\{\left(1-e^{-\frac{\gamma}{\bar{\gamma}_{RF}}}\right)^N + \left(1-e^{-\lambda\sqrt{\frac{\gamma}{\bar{\gamma}_{FSO}}}}\right) - \left(1-e^{-\frac{\gamma}{\bar{\gamma}_{RF}}}\right)^N\left(1-e^{-\lambda\sqrt{\frac{\gamma}{\bar{\gamma}_{FSO}}}}\right)\right\}d\gamma \quad (36)$$

Substituting the binomial expansion of $\left(1-e^{-\gamma/\bar{\gamma}_{RF}}\right)^N$ and Meijer-G equivalent of $e^{-\lambda\sqrt{\gamma/\bar{\gamma}_{FSO}}}$ as $\frac{1}{\sqrt{\pi}}G_{0,2}^{2,0}\left(\lambda^2\gamma/4\bar{\gamma}_{FSO}\Big|_{0,\frac{1}{2}}^{-}\right)$ and using [16, Eq. 07.34.03.1081.01] BER of DPSK modulation over negative exponential atmospheric turbulence equals with:

$$P_e = \frac{1}{2}\left\{1 - \frac{1}{\sqrt{\pi}}G_{1,2}^{2,1}\left(\frac{\lambda^2}{4\bar{\gamma}_{FSO}}\Big|_{0,\frac{1}{2}}^{0}\right) + \sum_{k=0}^{N}\binom{N}{k}(-1)^k\frac{1}{1+\frac{k}{\bar{\gamma}_{RF}}}\frac{1}{\sqrt{\pi}}G_{1,2}^{2,1}\left(\frac{\lambda^2}{4\bar{\gamma}_{FSO}\left(1+\frac{k}{\bar{\gamma}_{RF}}\right)}\Big|_{0,\frac{1}{2}}^{0}\right)\right\} \quad (37)$$

## 5 Numerical Results

In this section analytic and MATLAB simulation results for performance investigation of proposed hybrid FSO/RF system are compared. RF link fading has Rayleigh distribution and FSO link atmospheric turbulence in moderate to strong regime has Gamma-Gamma distribution with effect of pointing errors and in saturate regime has Negative Exponential distribution. Average received SNR of FSO and RF receivers are considered to be equal ($\bar{\gamma}_{FSO} = \bar{\gamma}_{RF} = \gamma_{avg}$). N is number of users. $\gamma_{th}$ is the proposed system outage threshold SNR. $\eta = 1$ and in fixed gain structure, C=1.

In Fig. 2 outage probability of proposed structure is plotted in terms of average SNR for moderate ($\alpha = 4, \beta = 1.9, \xi = 10.45$) and strong ($\alpha = 4.2, \beta = 1.4, \xi = 2.45$) regimes of Gamma-Gamma atmospheric turbulence with the effect of pointing error when $N = 2$ and $\gamma_{th} = 10dB$ for fixed and adaptive gain schemes. As can be seen, in $\gamma_{avg} = 30dB$, at different target $P_{out}$, difference of $\gamma_{avg}$ between moderate and strong regimes in adaptive gain scheme is less than $4dB$ and in fixed gain scheme is less than $3dB$; But by reducing $P_{out}$, this difference increases. Therefore, system performance of both fixed and adaptive gain schemes, at low $\gamma_{avg}$, is almost independent of the atmospheric turbulence. Thereby the proposed system is suitable for mobile communications. In mobile communication a small mobile battery, supplies transmitter power, therefore received $\gamma_{avg}$ will be low. As a result, the proposed structure does not need an adaptive processor and this reduces its power consumption, cost and complexity. In the case of fixed gain, $P_{out}$ is less than adaptive gain. Since in fixed gain, regardless of the channel conditions, amplification gain is chosen such that favorable performance can be achieved in all circumstances, but in adaptive gain, amplification gain changes with increase or decrease in atmospheric turbulence intensity. So fixed gain, despite more power consumption, has better performance but it is not economically affordable, because when channel condition is normal, most of the power is wasted.

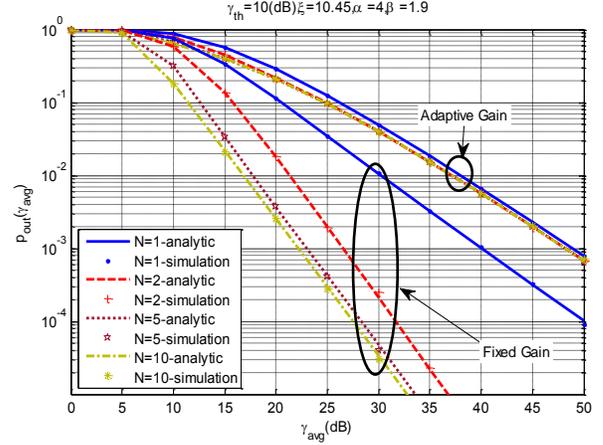

**Fig. 3:** Outage probability of proposed structure in terms of average SNR for different number of users (N) for moderate ($\alpha = 4, \beta = 1.9, \xi = 10.45$) regime of Gamma-Gamma atmospheric turbulence with the effect of pointing error when $\gamma_{th} = 10dB$, for fixed and adaptive gain schemes.

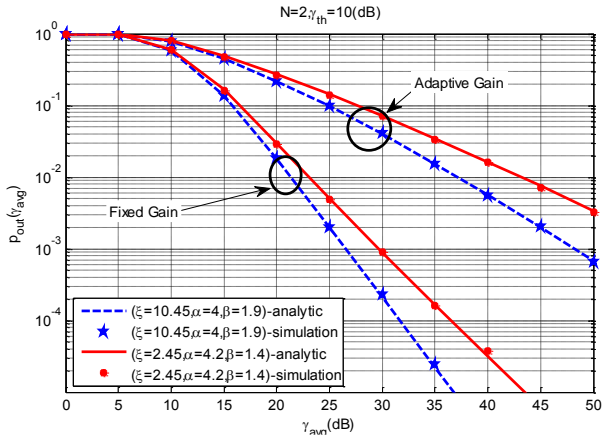

**Fig. 2:** Outage probability of proposed structure in terms of average SNR for moderate ($\alpha = 4, \beta = 1.9, \xi = 10.45$) and strong ($\alpha = 4.2, \beta = 1.4, \xi = 2.45$) regimes of Gamma-Gamma atmospheric turbulence with the effect of pointing error when $N = 2$ and $\gamma_{th} = 10dB$, for fixed and adaptive gain schemes.

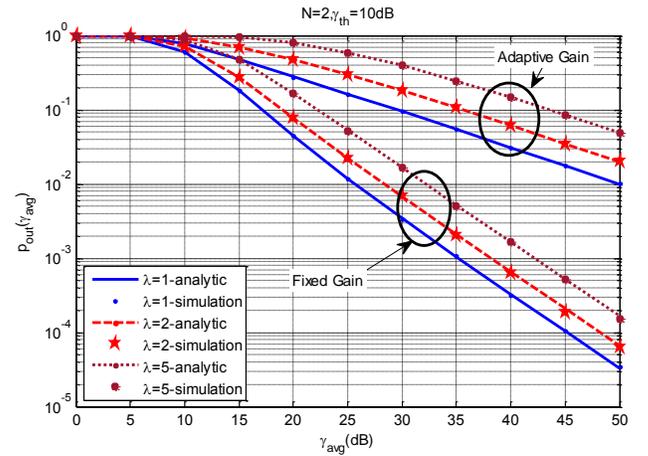

**Fig. 4:** Outage probability of the proposed structure is plotted in terms of average SNR for different variances ($1/\lambda^2$) of Negative Exponential atmospheric turbulence when N=2 and $\gamma_{th} = 10dB$ for fixed and adaptive gain schemes

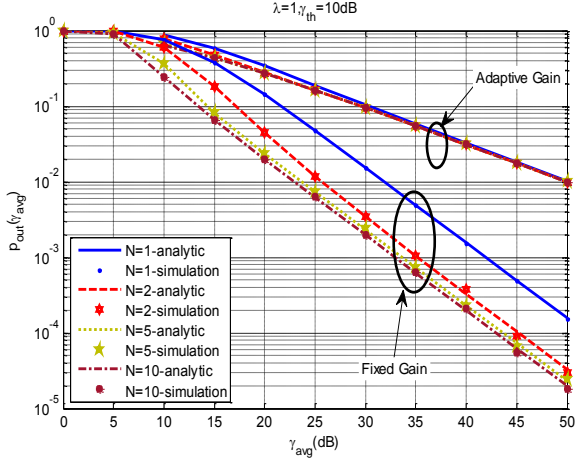

**Fig. 5:** Outage probability of proposed structure in terms of average SNR for different number of users for Negative Exponential atmospheric turbulences with unit variance when $\gamma_{th} = 10dB$, for fixed and adaptive gain schemes.

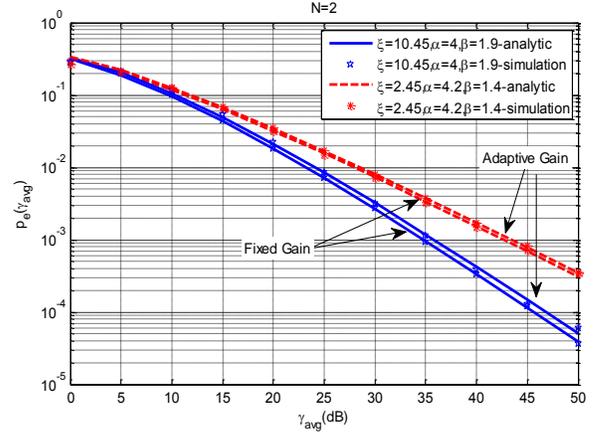

**Fig. 6:** Bit Error Rate of the proposed system in terms of average SNR for moderate ($\alpha = 4, \beta = 1.9, \xi = 10.45$) and strong ($\alpha = 4.2, \beta = 1.4, \xi = 2.45$) regimes of Gamma-Gamma atmospheric turbulence with the effect of pointing error when N = 2, for fixed and adaptive gain schemes.

In Fig. 3 outage probability of proposed structure is plotted in terms of average SNR for different number of users (N) in moderate ($\alpha = 4, \beta = 1.9, \xi = 10.45$) regime of Gamma-Gamma atmospheric turbulence with the effect of pointing error when $\gamma_{th} = 10dB$ for fixed and adaptive gain schemes. As can be seen, performance of the proposed system in adaptive gain scheme is not so much sensitive to the number of users, but in fixed gain, system is more sensitive to the number of users within the cell. Sensitivity of the system to number of users is due to selection done at the relay. In crowded cells, there are various independent fading paths, thereby probability that the SNR of all of received signals be below $\gamma_{th}$, is much smaller than the same probability for a low population cell. By increase of $\gamma_{avg}$, it is observed that the system performance in adaptive gain scheme is almost independent of the number of users and this is an advantage for the proposed system. This structure is suggested for areas where there is no direct RF link between user and base station. Actually these areas are low populated, so a system is suitable that performs independent of the number of users. In fixed gain, system performance in populated cell is much better, because its gain does not changes according to channel conditions and at different number of users a fixed gain is considered for relaying.

In Fig. 4 outage probability of the proposed structure is plotted in terms of average SNR for different variances $(1/\lambda^2)$ of Negative Exponential atmospheric turbulence when N=2 and $\gamma_{th} = 10dB$ for fixed and adaptive gain schemes. In this system, there are two users within the cell, as can be seen, at different target $P_{out}$, difference of $\gamma_{avg}$ between various variances of Negative Exponential atmospheric turbulence is the same. For example, when $P_{out} \leq 0.5$, difference of $\gamma_{avg}$ between cases of $\lambda = 1$ and $\lambda = 5$, in fixed gain is about $7dB$ and in adaptive gain is about $13dB$. This affects power consumption of system because at any $\gamma_{avg}$, adding a constant fraction of consumed power maintains system performance at various variances.

In Fig. 5 outage probability of proposed structure is plotted in terms of average SNR for different number of users (N) for Negative Exponential atmospheric turbulences with unit variance when $\gamma_{th} = 10dB$ for fixed and adaptive gain schemes. In the case of fixed gain, performance of the proposed system is sensitive to the number of users within the cell and the crowded cell has better performance because users face independent atmospheric turbulences and therefore the probability finding a user in a crowded cell with favorable SNR is much more than the same probability for a low population cell.

Bit Error Rate of the proposed system in terms of average SNR for moderate ($\alpha = 4, \beta = 1.9, \xi = 10.45$) and strong ($\alpha = 4.2, \beta = 1.4, \xi = 2.45$) regimes of Gamma-Gamma atmospheric turbulence with the effect of pointing error when N = 2, for fixed and adaptive gain schemes.

In Fig. 6 Bit Error Rate of the proposed system is plotted in terms of average SNR for moderate ($\alpha = 4, \beta = 1.9, \xi = 10.45$) and strong ($\alpha = 4.2, \beta = 1.4, \xi = 2.45$) regimes of Gamma-Gamma atmospheric turbulence with the effect of pointing error when N=2 for fixed and adaptive gain schemes. It can be seen when $\gamma_{avg} \leq 10dB$, link is virtually disrupted. Also, at low $\gamma_{avg}$, performance of the proposed system in different atmospheric turbulence regimes does not differ so much, but by increase of $\gamma_{avg}$ this difference increases. Behavior of the system in fixed gain and adaptive gain schemes, for both moderate and strong atmospheric turbulence regimes is almost the same when $\gamma_{avg}$ is low, i.e. the system acts independent of atmospheric turbulence regime. At low $\gamma_{avg}$, the noise effect is so much high but at high $\gamma_{avg}$, its effect becomes negligible.

In Fig. 7 Bit Error Rate of the proposed structure in terms of average SNR for different variances $(1/\lambda^2)$ of Negative Exponential atmospheric turbulence when N=2 for fixed and adaptive gain schemes. As can be seen, amplify and forward with fixed gain has better performance than adaptive gain. In adaptive gain scheme, relay gain changes according to channel condition, but in fixed gain structure, relay gain is always constant and it is always adjusted for the worst case condition. Thereby adaptive gain has less power consumption than fixed gain.

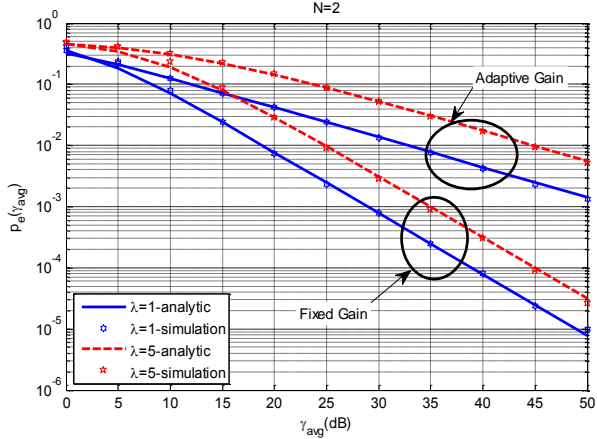 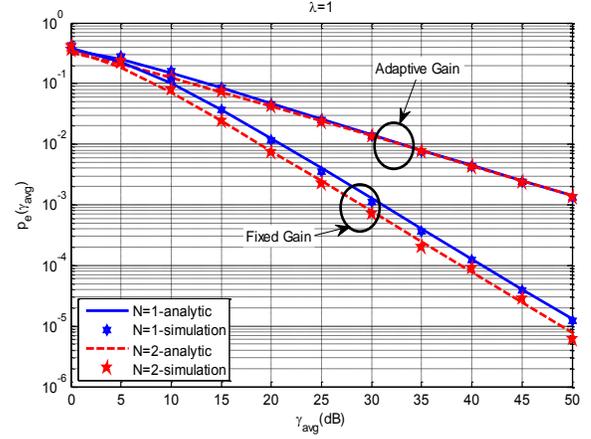

**Fig. 7:** Bit Error Rate of the proposed structure in terms of average SNR for different variances ($1/\lambda^2$) of Negative Exponential atmospheric turbulences when N=2 for fixed and adaptive gain schemes.

**Fig. 8:** Bit Error Rate of the proposed structure in terms of average SNR for different number of users (N) for Negative Exponential atmospheric turbulence with unit variance for fixed and adaptive gain schemes.

In Fig. 8 Bit Error Rate of the proposed structure is plotted in terms of average SNR for different number of users for Negative Exponential atmospheric turbulence with unit variance for fixed and adaptive gain schemes. It can be seen that in fixed gain structure, when only one user is added to the cell, at $P_e = 0.1$, $\gamma_{avg}$ increases about 2dB. In adaptive gain scheme at $\gamma_{avg} = 30 dB$, there is little difference in BER of the proposed system for different number of users, but since then system performs independent of number of users. Environments like seas mostly experience Negative Exponential atmospheric turbulence. Usually there is few number of users in these places, therefore a communication system is convenient for these areas which performs independent of number of users.

## 6 Conclusion

In this paper, a novel model for relay assisted hybrid FSO / RF communication system with series structure is presented. In this system a relay through a multiuser RF link, communicates with mobile users and through a FSO link communicates with base station. This structure is recommended especially when there is no RF link between mobile user and the base station. For the first time, closed form expressions are derived for BER and $P_{out}$ of the proposed system for both fixed and adaptive gain relay schemes and simulation results, verified accuracy of these expressions. It is shown that fixed gain, despite less complexity but because of more power consumption, has better performance. Adaptive gain relaying has low sensitivity to the number of users, but fixed gain is more sensitive. Overall, it can be concluded that in places with saturate atmospheric turbulence, because of the importance of power and the number of users, adaptive gain structure is suggested and in moderate to strong atmospheric turbulence, depending on power or system performance requirement, fixed or adaptive gain would be deployed.

## 7 References

1  Amirabadi M. A., Vakili V. T.: 'Performance analysis of a novel hybrid FSO / RF communication system', arXiv preprint arXiv:1802.07160, (2018)

2  Nebuloni R.: 'Empirical relationships between extinction coefficient and visibility in fog', Applied optics, 44(18), 3795-3804. (2005)

3  Amirabadi M. A., Vakili V. T.: 'A new optimization problem in FSO communication system', arXiv preprint arXiv:1802.07984, (2018)

4  Anees S., Bhatnagar M. R.: 'Performance of an Amplify-and-Forward Dual-Hop Asymmetric RF–FSO Communication System', Journal of Optical Communication Network 7(2), (2015)

5  Amirabadi M. A., Vakili V. T.: 'Performance Comparison of two novel Relay-Assisted Hybrid FSO/RF Communication Systems', arXiv preprint arXiv:1802.07335, (2018)

6  Mai V. V., Pham A. T.: 'Adaptive Multi-Rate Designs and Analysis for Hybrid FSO/RF Systems over Fading Channels', IEICE Transactions on Communications, E98-B(8) :1660-1671, (2015)

7  Amirabadi M. A., Vakili V. T.: 'A novel hybrid FSO/RF communication system with receive diversity', arXiv preprint arXiv:1802.07348, (2018)

8  Gappmair W., Nistazakis H. E.: 'Subcarrier PSK Performance in Terrestrial FSO Links Impaired by Gamma-Gamma Fading, Pointing Errors, and Phase Noise', Journal of Lightwave Technology 35(9), (2017)

9  Soleimani-Nasab E., Uysal M.: 'Generalized Performance Analysis of Mixed RF/FSO Cooperative Systems', IEEE Transactions on Wireless Communications 15(1), (2016)

10  Amirabadi M. A., Vakili V. T.: 'Performance analysis of hybrid FSO/RF communication systems with Alamouti Coding or Antenna Selection. arXiv preprint arXiv:1802.07286, (2018)

11  Lee E., Park J., Han D., Yoon G.: 'Performance Analysis of the Asymmetric Dual-Hop Relay Transmission With Mixed RF/FSO Links', IEEE Photonics Technology Letters 23(21), (2011)

12  Amirabadi M. A., Vakili V. T.: 'On the Performance of a CSI Assisted Dual-Hop Asymmetric FSO/RF Communication System over Gamma-Gamma atmospheric turbulence considering the effect of pointing error', International Congress on Science and Engineering, (2018)

13  Kong L., Xu X., Hanzo L., Zhang H., Zhao C.: 'Performance of a Free-Space-Optical Relay-Assisted Hybrid RF/FSO System in Generalized M-Distributed Channels', IEEE Photonics Journal 7(5), (2015)

14  Zedini E., Soury H., Alouini M. S.: 'On the Performance Analysis of Dual-Hop Mixed FSO/RF Systems', IEEE Transactions on Wireless Communications 15(5), (2016)

15  Gradshteyn I. S., Ryzhik I. M.: 'Table of Integrals, Series, and Products. Seventh Edition', Elsevier Inc. (2007)

16  http://functions.wolfram.com/HypergeometricFunctions/

17  Djordjevic G. T., Petkovic M. I., Cvetkovic A. M., Karagiannidis G. K.: 'Mixed RF/FSO Relaying with Outdated Channel State Information', IEEE Journal on Selected Areas in Communications PP(99), (2015)

18  Papoulis A., Pillai S. U.: 'Probability, Random Variables, and Stochastic Presses', McGraw-Hill, 4th edition, pp.194, (2002)